\documentclass
[pra,twocolumn,showpacs,preprintnumbers,amsmath,amssymb]{revtex4}%
\usepackage{graphicx}
\usepackage{dcolumn}
\usepackage{bm}
\usepackage{amsmath}
\usepackage{amsfonts}
\usepackage{amssymb}%
\begin{document}
\preprint{APS/123-QED}
\title{Shape oscillations in non-degenerate Bose gases - transition from the
collisionless to the hydrodynamic regime }
\author{Ch.~Buggle$^{1}$}
\author{P.~Pedri$^{2}$}
\author{W.~von~Klitzing$^{1}$}
\thanks{Present address: IESL - FORTH, Vassilika Vouton, 711 10 Heraklion, Greece.}
\author{J.T.M.~Walraven$^{1}$}
\affiliation{1: FOM Institute for Atomic and Molecular Physics\mbox{,} Kruislaan
407\mbox{,} 1098 SJ Amsterdam\mbox{,} The Netherlands and Van~der~Waals-Zeeman
Institute of the University of Amsterdam\mbox{,} Valckenierstraat
65/67\mbox{,} 1018 XE The Netherlands}
\affiliation{2: Institut f\"{u}r Theoretische Physik III Universit\"{a}t Stuttgart\mbox{,}
Pfaffenwaldring 57 V, 70550 Stuttgart\mbox{,} Germany}
\date{\today}

\begin{abstract}
We investigate collective oscillations of non-degenerate clouds of $^{87}$Rb
atoms as a function of density in an elongated magnetic trap. For the
low-lying $M\!=\!0$ monopole-quadrupole shape oscillation we measure the
oscillation frequencies and damping rates. At the highest densities the
mean-free-path is smaller than the axial dimension of the sample, which
corresponds to collisionally hydrodynamic conditions. This allows us to cover
the cross-over from the collisionless to the hydrodynamic regime. The
experimental results show good agreement with theory. We also analyze the
influence of trap anharmonicities on the oscillations in relation to observed
temperature dependencies of the dipole and quadrupole oscillation frequencies.
We present convenient expressions to quantify these effects.

\end{abstract}

\maketitle

\section{Introduction}

Collisional hydrodynamics has gradually become an important issue for the
understanding of experiments with dilute quantum gases. When the atomic
mean-free-path is smaller than the characteristic dimensions of typical
elongated atomic clouds, the gas properties depend on the local density field
and exhibit collisional hydrodynamics rather than the collisionless dynamics
of a nearly ideal gas \cite{Pitaevskii Stringari 2003,Lifschitz Pitaevski
1981}. For Bose gases and Bose-Fermi mixtures it is difficult to penetrate
deeply into this collisional hydrodynamic regime as three-body
molecule-formation will give rise to fast decay of the samples
\cite{Pitaevskii Stringari 2003}. Therefore, also the transition region
between collisionless and hydrodynamic conditions is of substantial practical importance.

The hydrodynamic flow of classical fluids was described as early as 1755 by
the equation of motion of Euler \cite{Landau Lifschitz 1987}. The opposite
limit of collisionless flow is equally well understood since the work of
Maxwell and Boltzmann and the investigation of rarefied gas dynamics around
the turn of the last century \cite{Lifschitz Pitaevski 1981}. The transition
regime between collisionless and hydrodynamic conditions deserves special
attention as the crossover behavior is often non-intuitive as was already
noted by Knudsen in 1908 \cite{Knudsen 1908}. With the availability of trapped
ultracold gases there is a renewed interest in the collisional hydrodynamics.
For non-degenerate quantum gases in harmonic traps the absence of the familiar
wall-boundary condition of zero hydrodynamic flow at the sample edges gives
rise to a very close phenomenological similarity with the superfluid
hydrodynamics of Bose-Einstein condensates \cite{Pitaevskii Stringari
2003,Kagan 1997,Griffin 1997, Nikuni1998}. Collisional hydrodynamics also has
to be considered in two-component Fermi gases near inter-component Feshbach
resonances, where the intercomponent scattering length is tuned to large
values in order to optimize thermalization \cite{Stringari 2004,O'Hara
2002,Regal 2003,Bourdel 2003}.

The onset of collisional hydrodynamics was first observed at MIT
in measurements of the damping and frequency shifts of the
low-lying $M=0$ quadrupole shape oscillation of cigar-shaped
samples of the $^{23}$Na quantum gas, just above the Bose-Einstein
transition temperature $T_{c}$ \cite{Stamper}. Similar results
were obtained at the ENS-Paris with clouds of metastable triplet
helium (He$^{\ast}$) \cite{Leduc}. A demonstration of the
collisional crossover was given at JILA by measuring, for varying
density, the damping of the center of mass oscillations of two
distinguishable clouds of $^{40}$K, passing in anti-phase
\cite{Gensemer,Toschi2004}. At AMOLF we showed how hydrodynamic
conditions affect the BEC-formation process in elongated samples
and can give rise to substantial shape oscillations of the
condensates being formed \cite{Shvarchuck 2002,Buggle04}. Further,
hydrodynamic conditions were shown to give rise to anisotropic
expansion of thermal Bose gases after switching off the confining
field, which has important consequences for time-of-flight
thermometry \cite{Shvarchuck 2003,Gerbier}. Hydrodynamic effects
were observed more pronouncedly in the expansions of two-component
Fermi gases tuned near an inter-component Feshbach resonance
\cite{O'Hara 2002,Regal 2003,Bourdel 2003}. Also the investigation
of the macroscopic dynamics of two-component Fermi gases in the
BCS-BEC transition region requires detailed understanding of the
hydrodynamics \cite{Greiner 2003,Jochim 2003,Zwierlein
2003,Cubizolles 2003}.

In this paper we study the crossover from collisionless to hydrodynamic
conditions in non-degenerate clouds of $^{87}$Rb by measuring both the
frequency shift and the damping of the low-lying $M=0$ quadrupole shape
oscillation as a function of density. In accordance with theory \cite{Kagan
1997,Griffin 1997, Nikuni1998}, the frequency shifts down from $2\omega_{z}$
in the collisionless regime to $1.55\omega_{z}$ for collisionally hydrodynamic
clouds, with $\omega_{z}$ the axial frequency of our trap. Most of the shift
occurs over a narrow range of densities around the crossover density, where
the mean-free-path becomes comparable to the axial size of the sample. At this
density also the strongest damping is observed. All our data were taken for
temperatures $T>2T_{c}$ to avoid precursor phenomena close to the BEC
transition \cite{Nikuni 2004}. Hence, although the collisions are quantum
(i.e. \textit{s}-wave), the gas is statistically classical. As we observed a
temperature dependence of the oscillation frequencies, we derive theoretical
expressions to include the influence of trap anharmonicities, which cause this
effect. These expressions allow numerical evaluation for regular potentials.
Further, they allow us to derive convenient analytic approximations that apply
to \emph{any} elongated Ioffe-Pritchard trap.

\section{Background}

\label{sec:Background} For quantum gases well-above the degeneracy
temperature, all oscillatory modes are solutions to the classical Boltzmann
equation \cite{Pitaevskii Stringari 2003,Chapman Cowling 1970}
\begin{equation}
\frac{\partial f}{\partial t}+\mathbf{v}\cdot\nabla_{\mathbf{r}}%
f+\frac{\mathbf{F}}{m}\cdot\nabla_{\mathbf{v}}f=I_{\mathrm{coll}%
}[f],\label{equ:Boltz}%
\end{equation}
where $f=f(t,\mathbf{r},\mathbf{v})$ is the phase-space distribution-function
with $\mathbf{{r=}}\{r_{j}\}=(x,y,z)$ and $\mathbf{{v=}}\{v_{j}\}$ the
position and momentum vectors, $m$ is the atomic mass and $\mathbf{F}%
(\mathbf{r})=-\nabla_{\mathbf{{r}}}U(\mathbf{r})$ the force of the
trapping potential $U(\mathbf{r})$; $I_{\mathrm{coll}}[f]$ is the
classical collisional integral. For the case of $s$-wave
collisions with energy-independent cross section $\sigma$ it takes
the form
\begin{equation}
I_{\mathrm{coll}}[f]=\frac{\sigma}{4\pi}\int
d\mathbf{v}_{1}d\Omega^{\prime
}\left\vert \mathbf{v}_{1}-\mathbf{v}\right\vert [f_{1}^{\prime}%
\,f^{\prime}-f_{1}\,f]
\end{equation}
and describes, for a given position and time, the effect of
elastic collisions between a pair of atoms with initial velocities
$\mathbf{v}$ and $\mathbf{v}_{1}$ and final velocities
$\mathbf{v}^{\prime}$ and $\mathbf{v}_{1}^{\prime}$. The solid
angle $\Omega^{\prime}$ gives the direction of the final relative
velocity with respect to the initial one.

For \textit{isotropic} harmonic traps the normal modes are multipoles of order
$(L;M)$ \cite{Griffin 1997}. Oscillations in the dipole mode $(L=1)$ are
commonly used for measuring trap frequencies by observing the motion of the
center of mass of trapped clouds as a function of time. In harmonic traps
these oscillations are undamped, since for any pair of atoms also their center
of mass oscillates at the trap frequency $\omega$. As noted in
ref.\,\cite{Griffin 1997, Nikuni1998}, Boltzmann obtained in 1897 the
surprising result that for isotropic harmonic traps also the monopole (or
`breathing') mode $(L=0)$ is undamped, oscillating at frequency $2\omega$,
\emph{independent} of the density. The next normal mode solutions are shape
oscillations $(L\geq2)$. In the hydrodynamic limit they are (like the dipole
mode) both irrotational and divergence-free \cite{Griffin 1997, Nikuni1998},
and therefore also undamped. They oscillate at frequency $\sqrt{L}\,\omega$
\cite{Note}. In the collisionless regime they are again undamped but oscillate
at frequency $L\,\omega$. This difference in frequency results in damping in
the transition regime \cite{Kavoulakis 1998} due to collisional relaxation
towards equilibrium.

For \emph{elongated} harmonic traps, with axial direction $z$ and radial
coordinate $\rho=(x^{2}+y^{2})^{1/2}$, we distinguish three $(L=1;M=0,\pm1)$
dipole modes, oscillating uncoupled and undamped at frequencies $\omega_{z}$
and $\omega_{\rho}$. In the hydrodynamic limit, the monopole mode is coupled
to the $(L=2;M=0)$ quadrupole mode. Decoupling in terms of irrotational
solutions yields \cite{Kagan 1997, Griffin 1997, Nikuni1998}
\begin{align}
\omega^{2}  &  =\frac{1}{3}[5\omega_{\rho}^{2}+4\omega_{z}^{2}%
\nonumber\label{eqn.QuadrupoleFrequencies}\\
&  \qquad\pm\sqrt{25\omega_{\rho}^{4}+16\omega_{z}^{4}-32\omega_{\rho}^{2}
\omega_{z}^{2}}].
\end{align}
In the experiment described in this paper, we study the low-lying $M=0$
coupled monopole-quadrupole mode, corresponding to the minus sign in
Eq.\,(\ref{eqn.QuadrupoleFrequencies}). In this mode the radial size
oscillates in anti-phase with the axial size. For shortness we will refer to
it as the `quadrupole' mode with frequency $\omega_{Q}$ in all regimes,
although in the collisionless limit the axial and radial motion decouple and
the overall behavior is to be considered as a superposition of `1D-breathing
modes', showing dephasing behavior. This dephasing can be avoided by exciting
a pure axial oscillation. As follows directly from
Eq.\,(\ref{eqn.QuadrupoleFrequencies}), in the limit of very elongated clouds
$\left(  \omega_{\rho}\gg\omega_{z}\right)  $ we have $\omega_{Q}=\sqrt
{12/5}\,\omega_{z}\approx1.55\,\omega_{z}$.

The transition regime is less obvious. Describing the oscillation
phenomenologically by $e^{-i\,\omega t}$ the crossover takes the form
\cite{Kavoulakis 1998}
\begin{equation}
\omega^{2}=\omega_{\text{cl}}^{2}+\frac{\omega_{\text{hd}}^{2}-\omega
_{\text{cl}}^{2}}{1-i\,\omega\tilde{\tau}}, \label{equ:frequ}%
\end{equation}
where $\omega=\omega^{\prime}+i\omega^{\prime\prime}$ is the complex
quadrupole frequency for a given thermal relaxation time $\tilde{\tau}$;
$\omega_{\text{hd}}$ and $\omega_{\text{cl}}$ are the (real) frequencies of
this mode in the hydrodynamic $(\omega^{\prime}\tilde{\tau}\ll1)$ and
collisionless $(\omega^{\prime}\tilde{\tau}\gg1)$ limits, respectively.

To have an intuitive picture of the solutions of Eq.\,(\ref{equ:frequ}) one
can separate the real and imaginary parts of $\omega$ and make the
identification $(\omega_{Q}=\omega^{\prime},$ $\Gamma=-\omega^{\prime\prime}%
)$. For $\Gamma/\omega_{Q}\ll1$ we can approximate the imaginary part of the
solution by the convenient form
\begin{equation}
\Gamma\simeq\frac{\tilde{\tau}}{2}\frac{\omega_{\text{hd}}^{2}-\omega
_{\text{cl}}^{2}}{1+\omega_{\text{cl}}^{2}\tilde{\tau}^{2}},
\label{equ:DampCrossOver}%
\end{equation}
which underestimates the maximum damping by $23\%$. The solution for the real
part of Eq.\,(\ref{equ:frequ}) can be heuristically, but fairly accurately
$(\pm0.3\%)$ described by
\begin{equation}
\omega_{Q}\simeq\omega_{\text{hd}}+\left(  \omega_{\text{cl}}-\omega
_{\text{hd}}\right)  \left(  2/\pi\right)  \arctan\!\left(  \omega_{\text{cl}
}^{2}\tilde{\tau}^{2}\right)  . \label{equ:FrequCrossOver}%
\end{equation}
We define the `cross-over point' as the point where maximum damping occurs and
the frequency is at the intermediate value $\omega_{Q}=1/2\,(\omega
_{cl}+\omega_{hd})$. From Eqs.\,(\ref{equ:DampCrossOver}) and
(\ref{equ:FrequCrossOver}) this is seen to occur at the relaxation time
$\tilde{\tau}=\tilde{\tau}_{0}$, where $2\omega_{z}\tilde{\tau}_{0}=1$.

Eq.\,(\ref{equ:frequ}) can be obtained from the Boltzmann equation in the
relaxation time approximation \cite{Guery-Odelin 1999,Al Khawaja 2000}, where
the collisional integral is replaced by
\begin{equation}
I_{\mathrm{coll}}[f]\simeq-\left(  f-f_{\mathrm{le}}\right)  /\tau.
\end{equation}
Here $\tau$ is the relaxation time and $f_{\mathrm{le}}=f_{\mathrm{le}
}(t,x,v)$ the local thermal distribution, which has an isotropic momentum
distribution \cite{Pedri 2003}. For harmonic traps one has~\cite{Guery-Odelin
1999}
\begin{equation}
\tau^{-1}=\frac{\sqrt{2}}{5}\tau_{c}^{-1},\quad\mathrm{where}\quad\tau_{c}
^{-1}=\sqrt{2}\,n_{0}\overline{v}_{th}\sigma\label{eqn.collisional time}%
\end{equation}
is the elastic collision rate at the trap center \cite{Chapman Cowling 1970},
with $\overline{v}_{th}=(8k_{B}T/\pi\,m)^{1/2}$ the mean thermal velocity at
temperature $T$, $n_{0}$ the central density and $\sigma$ the
elastic-scattering cross section.

To arrive at Eq.\,(\ref{equ:frequ}) the relaxation time has to be renormalized
by a factor that depends both on the cloud shape and the mode considered. For
the $M=0$ quadrupole mode in very elongated harmonic traps one finds
$\tilde{\tau}=6/5\,\tau$ \cite{Guery-Odelin 1999,Al Khawaja 2000}.

\section{Experiment}

In our experiment we typically load $10^{10}$ atoms from the $^{87}$Rb source
described in ref.\,\cite{Dieckmann 1998} into a magneto-optical trap. After an
optical molasses stage we optically pump the atoms into the fully stretched
($5S_{1/2},F=2,m_{F}=2$) hyperfine state and transfer the cloud into a
Ioffe-Pritchard trap with frequencies $\omega_{z}/2\pi=7$ Hz and $\omega
_{\rho}/2\pi=8$ Hz and central field $B_{0}=37$ Gauss. Any remaining
population in the $m_{F}=1$ magnetic sublevel is removed by gravitational sag.
Then, we radially compress the cloud, changing the trap parameters to
$\omega_{\rho}/2\pi=19$ Hz at $B_{0}=8$ G. After a thermalization time of
$100$ ms we add, in a linear ramp over $0.5$ ms, a magnetic field
$B_{m}=487\,$mG, rotating at a frequency of $\nu_{m}=7\,$kHz orthogonally to
the trap axis, using the approach described in ref.\,\cite{Tiecke}. This gives
rise to a Time-Averaged-Potential (TAP) field with offset $B_{0,m}
\equiv\left(  B_{m}^{2}+B_{0}^{2}\right)  ^{1/2}$ and frequencies
\begin{align}
\label{equ.TAP1}\omega_{\rho,m}  &  =
\omega_{\rho}\frac{(1+0.5\,b^{2})^{1/2}
}{(1+b^{2})^{3/4} },\\
\omega_{z,m}  &
=\omega_{z}\frac{1}{(1+b^{2})^{1/4}}\label{equ.TAP2},
\end{align}
where $b=B_{m}/B_{0}$ is the modulation depth \cite{Tiecke}.

We continue the compression to $\omega_{z,m}/2\pi=16.8$ Hz, $\omega_{\rho
,m}/2\pi=474$ Hz and $B_{0,m}=634$ mG ($B_{0}=406$ mG, $b=1.2$). Then, we cool
the sample by forced rf-evaporation to the final temperature of a few
microkelvin. After reducing the density to the desired level by laser
depletion \cite{Laser Depletion}, the sample is thermalized during plain
evaporation periods of up to $2.5$ s , which is sufficiently long even for our
lowest densities. We then raise the rf-shield energy by a factor of 7 to avoid
evaporation losses during the measurements.

It is important to note here that the harmonic range $\rho_{0,m}$ where
Eqs.\,(\ref{equ.TAP1}) and (\ref{equ.TAP2}) hold is proportional to the
amplitude of the rotating field: $\rho_{0,m}=B_{m}/\alpha$, where $\alpha$ is
the radial gradient of the Ioffe quadrupole field. For regions outside the
harmonic range the frequencies revert to the unmodulated ones
\cite{BuggleThesis}. This implies a \textit{minimum} required value for
$B_{m}$ to assure that the harmonic radius of the TAP field exceeds the
thermal size of the sample.

\subsection{Excitation of the quadrupole mode}

\begin{figure}[ptb]
\includegraphics[width=80mm]{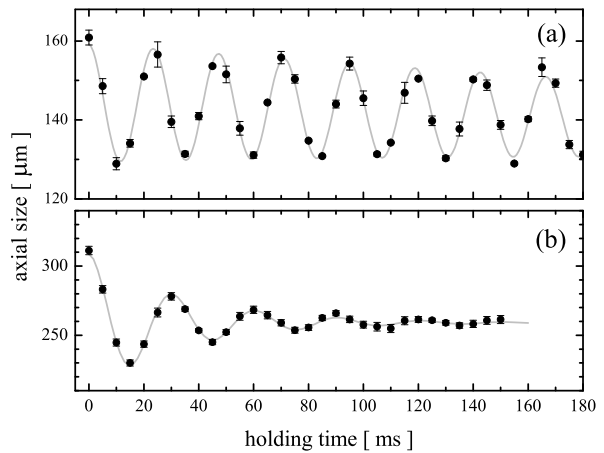}\caption{Typical quadrupole
oscillation traces for fitting frequency and damping; raw data acquired at a
temperature of $2\,\mu$K: a) Low density data ($n_{0}=1.7\times10^{11}%
$cm$^{-3}$) with fit function (grey line) resulting in $\omega_{Q}/\omega
_{z}=2.0$ and $\Gamma/\omega_{z}=0.01$. Each point represents the average of 3
absorption images.\newline b) High density data ($n_{0}=1.1\times10^{14}%
$cm$^{-3}$) with fit function (grey line) resulting in $\omega_{Q}/\omega
_{z}=1.6$ and $\Gamma/\omega_{z}=0.22$. Each point represents the average of
15 phase contrast images. The error bars represent the error in the mean.}%
\label{fig:Trace}%
\end{figure}To excite the quadrupole mode we remove the modulation field
${B}_{m}$ and observe the oscillation in a static potential. The advantage of
the TAP approach is the rapid switching between trap frequencies, which is
possible because both $B_{m}$ and $B_{0}$ are generated by trim coils. The
main currents of the Ioffe-Pritchard trap remain untouched. Further, this
approach offers definite knowledge of phase and amplitude. After transfer into
the static potential the cloud starts to oscillate inwards as a cosine
function with zero phase offset as can be seen in Fig.\,\ref{fig:Trace}.

As we remove the modulation, we simultaneously increase the central field to
$B_{0}=900$ mG in order to keep $\omega_{\rho}$ constant. The procedure is
done with a linear ramp of duration $\tau_{sw}=230~\mu$s. This is slow enough
to avoid switch-off depolarization and still much faster than the axial
oscillation time, $\,\omega_{z}\ll\,\tau_{sw}^{-1}\,\ll\,\omega_{Larmor}$.
Thus, after switching $\omega_{z}$, the gas finds itself diabatically in an
axially tighter potential. The axial trap frequency has increased to
$\omega_{z}/2\pi=21.1$ Hz, which changes the aspect ratio to $\omega_{\rho
}/\omega_{z}\approx23$ and puts us well into the elongated trap limit of
Eq.\,(\ref{eqn.QuadrupoleFrequencies}).

In this way we excite a pure axial oscillation, at least in the collisionless
limit. In the hydrodynamic limit, in principle both the low-lying and the
high-lying monopole-quadrupole modes could be excited. However, since even at
our highest densities, \emph{radially} we remain in the collisionless regime,
the high-lying mode cannot be excited due to the lack of coupling.

If the extent of the cloud prepared in the TAP-modulated magnetic field
reaches significantly beyond $\rho_{0,m}$, its density profile deviates from a
Gaussian. Transferring that distribution in the described way into the static
magnetic field leads to excitation of higher modes. If additionally the
thermal size of the cloud also exceeds the harmonic range of the static
potential (which is \textit{not} related to $\rho_{0,m}$), both
anharmonicities will add to the excitation of higher modes. However, these
modes oscillate at much higher frequencies and damp accordingly faster than
the quadrupole mode under investigation.

For our highest-density samples, together with the condition $T\gtrsim2T_{c}$,
we have $\rho_{0,m}\gtrsim10$ $\mu$m, which implies a required TAP amplitude
$B_{m}>350\,$mG. The value of $B_{m}=487\,$mG, used in the experiment,
represents our technical limit, and corresponds to the harmonic $1/e$-size of
a thermal cloud at a temperature of $9\,\mu$K. To assure that fitted values
for frequency and damping are unaffected by higher modes, we neglect the first
two cycles of oscillation traces acquired at temperatures above $7\,\mu$K, and
the first cycle for traces acquired above $4\,\mu$K. For lower temperatures
also the first cycle is analyzed. Note that the precise reproducibility of the
starting phase of our oscillations allows this procedure without degrading the
quality of the fits.

\subsection{Description of the trapping field}

During the observation of the quadrupole oscillation the cloud resides in a
potential given by $U(\mathbf{{r}})=\mu_{B}[B(\mathbf{{r}})-B_{0}]+mgy$ for
the chosen Zeeman-level in this experiment. Here $\mu_{B}$ is the Bohr
magneton and $g$ the gravity acceleration along the vertical direction
($y$-direction). For elongated Ioffe-Pritchard traps the modulus of the
trapping field $B(\mathbf{{r}})$ is accurately described by \cite{Bergeman
1987, Surkov 1994}
\begin{equation}
\label{eqn.TrapField}B(x,y,z)=\sqrt{(B_{0}+\beta z^{2})^{2}+\alpha^{2}
(x^{2}+y^{2})+4\alpha\beta xyz},
\end{equation}
where $B_{0}=0.9$ G and $\alpha=353$ G/cm are defined above and $2\beta=274$
G/cm$^{2}$ is the axial curvature. {To our knowledge $\alpha$ and $\beta$ were
constant throughout the measurements to within }${0.1\%}$; {$B_{0}$ was
monitored to be constant to within }${1\%}$. Expanding
Eq.\,(\ref{eqn.TrapField}) around the trap center and keeping the leading
non-linearities \cite{Surkov 1994}, the potential can be written as
\begin{align}
\label{eqn.TrapPotential}U(\rho,z)  &  = \frac{1}{2}m[\omega_{z}^{2}
z^{2}(1-\frac{1}{2}\rho^{2}/\rho_{0}^{2})+\nonumber\\
&  \qquad+\omega_{\rho}^{2}\rho^{2}(1-\frac{1}{4}\rho^{2} /\rho_{0}
^{2})]+\cdots,
\end{align}
where $m\omega_{z}^{2}=2\mu_{B}\beta$, $m\omega_{\rho}^{2}=\mu_{B}\alpha
^{2}/B_{0}$ and $\rho_{0}=B_{0}/\alpha=25$ $\mu$m the harmonic radius
\cite{Note on Gravity}.

\subsection{Detection procedure}

Two imaging methods are used to observe the oscillations. For our
highest-density samples we use phase-contrast imaging with red-detuned light.
For densities $n_{0}>5\times10^{13}$ cm$^{-3}$ a proper contrast is obtained
at a detuning of $-3$ GHz, where the detection is essentially non-destructive
\cite{Detection}. This allows us to register the oscillations in a sequence of
$31$ images at $5$ ms intervals, taking advantage of the fast `kinetics'
imaging mode of our camera \cite{CamType}. For lower densities the phase
contrast method cannot be used because, at the (smaller) detunings required to
maintain adequate phase contrast, photoassociation losses disturb the
measurements \cite{Note Photoassociation losses}.

For densities $n_{0}<5\times10^{13}$ cm$^{-3}$, we used repetitive absorption
imaging on the $(5S_{1/2},F=2)\leftrightarrow(5P_{3/2},F=3)$ transition
($D_{2}$-line) \cite{Note on detection}, varying the holding time of the cloud
after excitation of the oscillation. The images were taken \textit{in situ},
just before releasing the cloud from the trap \cite{Note 1}. We apply the
usual method of background subtraction and level-normalization to process the
images \cite{Ketterle 1999,Dieckmann 2001}. To retrieve the column density
profile $n_{2}(y,z)$ and the axial and radial Gaussian $1/e$-sizes $L_{e}$ and
$R_{e}$, we fit a 2-dimensional Gaussian expression to the optical thickness
distribution of our images. The central density follows with $n_{0}%
=n_{2}(0,0)/\sqrt{\pi R_{e} ^{2}}$ and, with Eq.\,(\ref{eqn.collisional time}%
), the relaxation time can be expressed as
\begin{equation}
\omega_{z}\tilde{\tau}=\frac{3}{2}\,\frac{\omega_{z}}{\omega_{\rho}}
\,\frac{\pi}{n_{2}(0,0)\sigma}. \label{eqn.OmegaTau}%
\end{equation}
Note that this expression does not depend explicitly on the gas temperature.
The collision cross section is $\sigma=8\pi a^{2}$ in the zero temperature
limit and is calculated with the value $a=98.98(4)a_{0}$ for the $s$-wave
scattering length \cite{Kempen 2002}.

To acquire sufficient statistics, at least $30$ images are taken to retrieve
one oscillation trace for a given density and each trace is acquired at least
$3$ times. Because the crossover happens over a narrow range of densities,
great care was taken to reproduce the initial conditions from shot to shot.
This is done by adjusting the density using laser depletion in a feedback loop
with the experimental result of the previous shot \cite{Laser Depletion}.
Although this procedure increases the shot to shot fluctuations, long-term
drift is virtually eliminated. With this procedure the atom number could be
long-term stabilized within a standard deviation of better than $1\%$. By
fitting the expression for an exponentially damped cosine function to the
trace (see Fig.\,\ref{fig:Trace}), we retrieve the experimental values for the
frequency $\omega_{Q}$ and damping rate $\Gamma$ of the quadrupole mode for
the selected density.

\subsection{Accuracy of density and temperature determination}

The \textit{absolute} accuracy of $n_{2}(0,0)$ is estimated to be $\sim30\%$
\cite{Note on linewidth}. The phase contrast images are calibrated against
absorption images of expanded clouds taken $15$ ms after release from the trap
at zero detuning. This procedure presumes the conservation of atom number
during the expansion.

In our analysis we account to leading order for the corrections associated
with trap anharmonicities. For temperatures much lower than the harmonic
temperature $T_{0}=\mu_{B}B_{0}/k_{B}=60$\,$\mu$K,
Eq.\,(\ref{eqn.TrapPotential}) becomes sufficiently accurate to describe the
cloud shape. In this limit the column density on the trap axis (to leading
order in the $x$-integration) can be expressed for $z^{2}\ll2k_{B}
T/m\omega_{z}^{2}$ as
\begin{equation}
n_{2}(0,z)\simeq n_{2}(0,0)\exp\!\left(  -\frac{m\omega_{z}^{2}z^{2}}{2k_{B}T}
(1-\frac{1}{2}T/T_{0})\right)  , \label{eqn.ColumnDensity}%
\end{equation}
where $T/T_{0}=\left\langle x^{2}\right\rangle /\rho_{0}^{2}=$ $k_{B}T/\mu
_{B}B_{0}$ with $\left\langle x^{2}\right\rangle =k_{B}T/m\omega_{\rho}^{2}$
the variance of the thermal distribution of the cloud along the $x$-axis in
the harmonic limit.

From Eq.\,(\ref{eqn.ColumnDensity}) we estimate the $1/e$-axial-size $L_{e}$
that will be obtained by fitting a Gaussian to the axial column density
profile of the cloud, $L_{e}=L/(1-\frac{1}{4}T/T_{0})$ with $L$ defined by
$L^{2}=2k_{B}T/m\omega_{z}^{2}$ \cite{Note 2D versus 3D fitting}. The
temperature follows with the expression
\begin{equation}
\label{eqn.CloudTemperature}k_{B}T\simeq\frac{1}{2}m\omega_{z}^{2}L_{e}
^{2}(1-\frac{1}{2}T/T_{0} ).
\end{equation}
Hence, for a temperature of $6$ $\mu$K the harmonic approximation
overestimates the temperature by $\sim5\%$. The correction in the central
column density is smaller. Numerically we established that the fit of a 2D
Gaussian underestimates the central column density by $\sim1.4\%$ at
$T/T_{0}=0.1$. As these corrections are small, there is no need to go beyond
the leading order of anharmonic correction to retrieve these quantities. For
measuring oscillation frequencies the situation is different because these can
be measured to high precision.

Mean field broadening of the distribution is small \cite{Guery02}. Calculating
the variance $\left\langle z^{2}\right\rangle =\frac{1}{2}L^{2} $ using the
recursive expression for the density to first order in mean field
$U_{\text{mf}}(\mathbf{r})=2v_{0}n(\mathbf{r})$, leads for $T\ll T_{0}$ to
\begin{equation}
\frac{1}{2}m\omega_{z}^{2}L^{2}\simeq k_{B}T+E_{\text{mf}}, \label{T0Emf}%
\end{equation}
where $E_{\text{mf}}=v_{0}\int n^{2}(\mathbf{r})d\mathbf{r}/\int
n(\mathbf{r})d\mathbf{r}=v_{0} n_{0}/\sqrt{8}$ is the trap averaged
interaction energy with $v_{0}=(4\pi\hbar^{2}/m)a$ the interaction coupling
constant \cite{Pitaevskii Stringari 2003}. Equivalently, treating the mean
field as an effective potential we may write
\begin{equation}
k_{B}T=\frac{1}{2}m\omega_{z}^{2}L^{2}(1-\xi), \label{T0}%
\end{equation}
where $\xi=E_{\text{mf}}/(k_{B}T+E_{\text{mf}})$ is the mean field correction
constant. For the data point with the highest mean field ($n_{0}%
=1.1\times10^{14}$cm$^{-3}$, $T=2\,\mu$K) we calculate $\xi=0.007$. Therefore,
mean field corrections are at least one order of magnitude smaller that the
anharmonic corrections and are discarded in this paper.

\section{Anharmonic frequency shifts}

\label{sec:nonlinear} As we operate at temperatures well above $T_{c}$, we pay
special attention to the issue of trap anharmonicities. We follow the path of
argumentation as presented in \cite{Guery-Odelin 1999,Pedri 2003} to derive
expressions for the anharmonic shifts. These are both temperature and mode
dependent and can also depend on the density. The expressions are suitable for
numerical evaluation provided the first and second spatial derivatives of the
trapping potential are known.

To describe the dynamical evolution of an observable $\chi=\chi(\mathbf{r},
\mathbf{v})$ it is multiplied by Eq.\,(\ref{equ:Boltz}) and averaged over the
phase space
\begin{equation}
\frac{d}{dt}\langle\chi\rangle-\langle\mathbf{v}\cdot\nabla_{\mathbf{r}}
\chi\rangle-\frac{1}{m}\langle\mathbf{F}\cdot\nabla_{\mathbf{v}}\chi
\rangle=-\frac{\langle\chi\rangle-\langle\chi\rangle_{\mathrm{le}}}{\tau},
\end{equation}
where
\begin{equation}
\langle\chi\rangle(t)=\frac{1}{N}\int\chi(\mathbf{r},\mathbf{v})f(t,\mathbf{r}
,\mathbf{v})d^{3}rd^{3}v, \label{equ:AvgOnPhaseSpace}%
\end{equation}
with $N$ the number of atoms. By choosing the correct set of observables, it
is possible to obtain a closed set of equations that describes the dynamics of
these observables.

\subsection{Dipole mode $(L=1)$}

\begin{figure}[ptb]
\includegraphics[width=80mm]{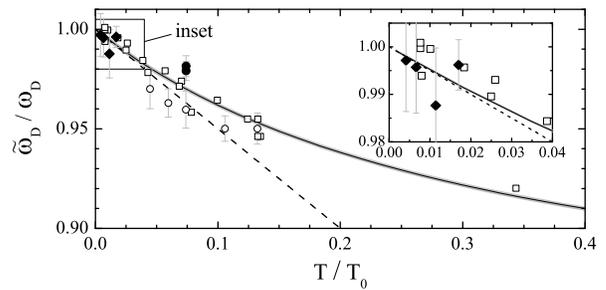}\caption{Scaled frequencies
$\tilde{\omega}_{D}/\omega_{D}$ for the axial dipole mode versus temperature.
The grey line corresponds to the evaluation of Eq.\,(\ref{equ:dipole}). The
leading slope (dashed line) is given by Eq.\,(\ref{eqn.LeadingShiftDipole}).
Open circles: phase contrast measurements; closed circles: absorption imaging
measurements, both acquired with thermal clouds. Diamonds: measurements with
Bose-Einstein condensates, using absorption imaging. Open squares: results
acquired at ENS-Paris \cite{Jeremi}.}%
\label{fig:COMFrequ}%
\end{figure}To investigate the effect of trap anharmonicities on the dipole
mode oscillation, we make the Ansatz
\begin{equation}
f(t,\mathbf{r},\mathbf{v})=f_{0}(r_{i}-a_{i},v_{i}-\dot{a}_{i}),
\end{equation}
where $f_{0}(\mathbf{r},\mathbf{v})=\mathcal{C}\exp\!\left(  -\left(
mv^{2}/2+U(\mathbf{r})\right)  /k_{B}T\right)  $ { is the equilibrium
distribution function }with $\mathcal{C}$ the normalization factor and
$a_{i}=a_{i}(t)$. We choose $\chi=v_{i}$ and obtain the following set of
equations {\setlength\arraycolsep{2pt}
\begin{align}
\frac{d}{dt}\langle v_{i}\rangle- \frac{1}{m}\langle F_{i}(\mathbf{r} )\rangle
&  = 0, \quad\mathrm{where}\\
\phantom{M}\nonumber\\
\langle v_{i}\rangle=\dot{a}_{i}\quad\mathrm{and}\quad\langle F(\mathbf{r}%
)\rangle &  = \langle F(\mathbf{r}+\mathbf{a})\rangle_{0}.
\end{align}
} Analogously to Eq.\,(\ref{equ:AvgOnPhaseSpace}) we denote with $\langle
\chi\rangle_{0}$ the average on the phase space using the equilibrium
distribution $f_{0}(\mathbf{r},\mathbf{v})$. Expanding up to first order
around the equilibrium position $a_{i}=0$, we obtain
\begin{equation}
\ddot{a}_{i}+\frac{1}{m}\sum_{j}\langle U_{ij}^{\prime\prime}\rangle_{0}
a_{j}=0,
\end{equation}
where $U_{ij}^{\prime\prime}=\frac{\partial^{2}U}{\partial r_{i}\partial
r_{j}}$. Restricting ourselves to potentials with $\langle U_{ij}
^{\prime\prime}\rangle_{0}=0$ for $i\neq j$, we obtain for the effective
frequencies of the dipole modes
\begin{equation}
\label{equ:dipole}\tilde{\omega}_{iD}^{2}=\frac{1}{m}\langle U_{ii}
^{\prime\prime}\rangle_{0}.
\end{equation}
Substituting Eq.\,(\ref{eqn.TrapPotential}) for the potential into
Eq.\,(\ref{equ:dipole}) we obtain for the leading anharmonic shift in the
$z$-direction
\begin{equation}
\label{eqn.LeadingShiftDipole}\tilde{\omega}_{zD}\simeq\omega_{z}(1-\frac
{1}{2}T/T_{0} ).
\end{equation}
This expression is shown as the dashed line in Fig.\,\ref{fig:COMFrequ}. The
integral in Eq.\,(\ref{equ:dipole}) is readily evaluated numerically using
Eq.\,(\ref{eqn.TrapField}) and requires as input parameters only the values
for $\alpha$, $\beta$ and $B_{0}$. The resulting curve is shown as the solid
line in Fig.\,\ref{fig:COMFrequ}. The curve follows the trend of our
measurements of center-of-mass oscillations as well as data obtained in Paris
\cite{Jeremi}.

The zero temperature limit of $\tilde{\omega}_{zD}$ is largely fixed by
measurements with Bose-Einstein condensates, which reproduced within $1\%$
over a period of one year. Its value is used to calibrate $\omega_{z}$ and the
related $\beta$-coefficient. We have no explanation for the remaining
deviations for the points taken with thermal samples at higher temperatures
\cite{Repeating}. We cannot trace them back to insufficient mechanical or
electronic stability of our trap. Non-exponential contributions to the damping
may account for a systematic error in the frequency, but should be less than
$1\%$. We speculate that possibly the temperature determinations of the phase
contrast measurements could be affected by a molecular contribution to the
phase contrast, which tends to narrow-down the distribution and results in an
underestimated value for the temperature. This results from the distribution
of pairs, that can photoassociate, which is proportional to the square of the
atomic density.

\subsection{{Surface modes }${(L=0,L=2)}$}

In order to calculate the anharmonic shifts for the breathing and the two
quadrupole modes we make the Ansatz
\begin{equation}
f(t,\mathbf{r},\mathbf{v})=\frac{1}{\prod_{j}b_{j}\sqrt{\theta_{j}}}
\;f_{0}\!\!\left(  \frac{r_{i}}{b_{i}},\frac{(v_{i}-(\dot{b}_{i}/b_{i})r_{i}
)}{\theta_{i}^{1/2}}\right)  ,
\end{equation}
where $b_{i}$ and $\theta_{i}$ are time-dependent variables. The parameters
$b_{i}$ take into account shape deformation of the density cloud whereas the
parameters $\theta_{i}$ allow an anisotropic momentum distribution which is
crucial to calculate the correct frequencies. We choose $\chi=v_{i}r_{i}$ and
obtain the following set of equations
\begin{equation}
\ddot{b}_{i}\langle r_{i}^{2}\rangle_{0}-\frac{\theta_{i}}{b_{i}}\langle
v_{i}^{2}\rangle_{0}-\frac{1}{m}\langle F_{i}(b_{j}r_{j})r_{i}\rangle_{0}=0.
\end{equation}
We impose the stationary solution and find the relation
\begin{equation}
\langle v_{i}^{2}\rangle_{0}=-\frac{1}{m}\langle F_{i}(r_{j})r_{i}\rangle_{0}.
\end{equation}
Then choosing $\chi=(v_{i}-(\dot{b}_{i}/b_{i})r_{i})^{{2}}$ yields
\begin{equation}
\dot{\theta}_{i}+2\frac{\dot{b}_{i}}{b_{i}}\theta_{i}=-\frac{\theta_{i}
-\bar{\theta}}{\tau},
\end{equation}
where $\bar{\theta}=(\sum_{i}\theta_{i})/3$.

Let us now focus our attention on two extreme regimes.

\subsubsection{Collisionless limit}

\begin{figure}[ptb]
\includegraphics[width=80mm]{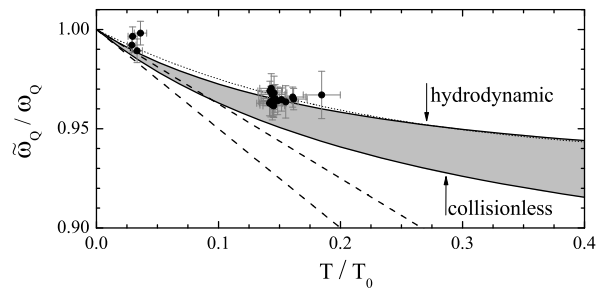}\caption{Scaled frequencies
$\tilde{\omega}_{Q}/\omega_{Q}$ for the quadrupole mode versus temperature.
The grey sector corresponds to Eq.\,(\ref{eqn.Quadrupole}) evaluated for the
cross-over regime (see Sec.\ref{sec.CrossOverRegime}) with the collisionless
and hydrodynamic limits indicated. The dashed lines show the leading slopes
given by Eqs.\,(\ref{eqn.LeadingShiftCollisionless}) and
(\ref{eqn.LeadingShiftHydrodynamic}). All data points correspond to
collisionless conditions. The dotted line is used to scale all quadrupole
frequency data to zero temperature.}%
\label{fig:QUADFrequ}%
\end{figure}In the collisionless limit ($\tau\rightarrow\infty$) we obtain the
relation
\begin{equation}
\theta_{i}=\frac{1}{b_{i}^{2}},
\end{equation}
and finally,
\begin{equation}
\label{equ:DiffEquC}\ddot{b}_{i}+\frac{1}{b_{i}^{3}}\frac{\langle F_{i}
(r_{j})r_{i}\rangle_{0} }{m\langle r_{i}^{2}\rangle_{0}}-\frac{\langle
F_{i}(b_{j}r_{j})r_{i} \rangle_{0}}{m\langle r_{i}^{2}\rangle_{0}}=0.
\end{equation}

Linearizing these equations and looking for solution of the form $e^{-i\omega
t}$ we obtain the three frequencies. In order to do this explicitly we define
the quantities
\begin{equation}
A_{ii}=\frac{3}{m}\frac{\langle r_{i}U_{i}^{\prime}\rangle_{0}}{\langle
r_{i}^{2}\rangle_{0}},
\end{equation}
where $U_{i}^{\prime}=\frac{\partial U}{\partial r_{i}}$, $A_{ij}=0$ for
$i\neq j$ and
\begin{equation}
B_{ij}=\frac{1}{m}\frac{\langle r_{i}r_{j}U_{ij}^{\prime\prime}\rangle_{0}
}{\langle r_{i}^{2}\rangle_{0}},
\end{equation}
where $U_{ij}^{\prime\prime}=\frac{\partial^{2}U}{\partial r_{i}\partial
r_{j}}$; note that, in general, $B_{ij}\neq B_{ji}$. We have to solve
\begin{equation}
\label{eqn.Determinant}|A+B-\tilde{\omega}^{2}I|=0,
\end{equation}
where $I$ is the identity-matrix, in order to obtain the frequencies. For the
quadrupole modes with $M=\pm2$ we find
\begin{equation}
\tilde{\omega}_{Q2}^{\phantom{Q2}2}=A_{xx}-A_{xy}+B_{xx}-B_{xy} \label{eqn.Q2}%
\end{equation}
whereas for the modes with $M=0$ we have
\begin{align}
\label{eqn.Breathing}\tilde{\omega}_{\mathrm{B}}^{\phantom{B}2} &
=
\mu+\sqrt{\mu^{2}-\Delta}\\
\tilde{\omega}_{Q}^{\phantom{Q}2}  &  =\mu-\sqrt{\mu^{2}-\Delta} ,
\label{eqn.Quadrupole}
\end{align} where
\begin{align}
\label{eqn.Mu}\mu &  =\left(  A_{xx}+A_{xy}+A_{zz}+B_{xx}+B_{xy}%
+B_{zz}\right)  /2 \phantom{M}\\
\Delta &  =(A_{xx}+A_{xy}+B_{xx}+B_{xy})(A_{zz}+B_{zz})\nonumber\\
&  \quad-2(A_{xz}+B_{xz})(A_{zx}+B_{zx}). \label{eqn.Delta}%
\end{align}
Here we used {$\langle U_{ij}^{\prime\prime}\rangle_{0}=0$ for $i\neq j$.} An
analytic approximation for the leading anharmonic shift of the $M=0$
quadrupole mode is obtained by substituting Eq.\,(\ref{eqn.TrapPotential}) for
the trap potential,
\begin{equation}
\label{eqn.LeadingShiftCollisionless}\tilde{\omega}_{Q}\simeq\omega
_{Q}(1-\frac{1}{2}T/T_{0} ),
\end{equation}
where $\omega_{Q}$ is the frequency in the harmonic limit. Note that at this
level of approximation the relative shift coincides with that of the dipole
mode. The result of the numerical averages based on Eq.\,(\ref{eqn.TrapField})
is shown as the lower solid line in Fig.\,\ref{fig:QUADFrequ}. The lower
dashed line corresponds to the leading shift given by
Eq.\,(\ref{eqn.LeadingShiftCollisionless}).

Comparison with the experimental points in Fig.\,\ref{fig:QUADFrequ} shows
agreement as far as the trend of the shift is concerned but a systematic
deviation for the slope. This discrepancy can be eliminated by presuming that
our axial trap frequency $\omega_{z}$ is underestimated by $1.5\%$. However,
such a correction cannot be justified on the basis of the limited set of data
for the dipole mode \cite{Repeating}.

\subsubsection{Hydrodynamic limit}

In the hydrodynamic regime $(\tau\rightarrow0)$ the local equilibrium is
always maintained, which implies that $\theta_{i}=\bar{\theta}$. In this case
we obtain the relation
\begin{equation}
\theta_{i}=\bar{\theta}=\frac{1}{\prod_{i}b_{i}^{2/3}},
\end{equation}
and therefore
\begin{equation}
\ddot{b}_{i}+\frac{1}{b_{i}(\prod_{j}b_{j})^{2/3}}\frac{\langle F_{i}
(r_{j})r_{i}\rangle_{0}}{m\langle r_{i}^{2}\rangle_{0}}-\frac{\langle
F_{i}(b_{j}r_{j})r_{i}\rangle_{0}}{m\langle r_{i}^{2}\rangle_{0} }=0.
\label{equ:DiffEquH}%
\end{equation}
By linearizing around the equilibrium we find the frequencies for the $M=\pm2$
modes and the two $M=0$ monopole-quadrupole modes. In this case we have to
define the $A_{ij}$ matrix as
\begin{equation}
A_{ii}=\frac{5}{3m}\frac{\langle
r_{i}U_{i}^{\prime}\rangle_{0}}{\langle
r_{i}^{2}\rangle_{0}}\text{ ~and ~}A_{ij}=\frac{2}{5}A_{ii}.
\end{equation}
Note that $A_{ij}$ does not depend on $j$. The matrix $B_{ij}$ is the same as
in the collisionless case. Solving the determinant Eq.\,(\ref{eqn.Determinant}%
) leads again to Eqs.\,(\ref{eqn.Q2}), (\ref{eqn.Breathing}) and
(\ref{eqn.Quadrupole}) for the frequencies and
Eqs.\,(\ref{eqn.Mu}) and (\ref{eqn.Delta}) for $\mu$ and $\Delta$.
Only the expressions for the matrix elements $A_{ij}$ have
changed. Substituting Eq.\,(\ref{eqn.TrapPotential}) for the trap
potential we find for the leading anharmonic shift of the
hydrodynamic $M=0$ quadrupole mode
\begin{equation}
\label{eqn.LeadingShiftHydrodynamic}\tilde{\omega}_{Q}\simeq\omega_{Q}
(1-\frac{3}{8}T/T_{0} ),
\end{equation}
which has a slightly weaker slope than in the collisionless case. The result
of the numerical averages based on Eq.\,(\ref{eqn.TrapField}) are shown as the
upper solid line in Fig.\,\ref{fig:QUADFrequ}. The upper dashed line
corresponds to the leading shift given by
Eq.\,(\ref{eqn.LeadingShiftHydrodynamic}).

A comparison with experiment requires densities $n_{0}>4\times10^{14}%
\,$cm$^{-3}$ at a temperature $T=4$ $\mu$K, to have $2\omega_{z}\tilde{\tau
}<0.1$, which is about three times our maximum density. At our highest density
of $n_{0}=1.3\times10^{14}$cm$^{-3}$, we calculate a 3-body decay rate of
$\dot{N}/N=2\sqrt{3}\,L\,n_{0}^{2}\approx1$ s$^{-1}$, with $L\!=\!
1.8(5)\times10^{-29} $cm$^{6}$s$^{-1}$ the three-body rate constant in the
Bose-condensed state \cite{Soeding1999}. At a three times higher density, the
decay rate renders the acquisition of data at approximately constant density
impossible for $^{87}$Rb.

\subsubsection{Crossover regime}

\label{sec.CrossOverRegime}

In the cross-over region, the same approach can be used, but after
linearizing, one has to look for solutions of the form $e^{-i\,\omega t}$ with
a complex $\omega$. For the $M=0$ modes this leads to the equation
\begin{equation}
\label{equ:CrossOverFrequVsTemp}\left(  C[\omega]-\frac{i}{\tau}
D[\omega]\right)  \left(  E[\omega]-\frac{i}{\tau}F[\omega]\right)  =0,
\end{equation}
where $C[\omega]=\omega(\omega^{2}-\omega_{\mathrm{\mathrm{cl:B}}}^{2}
)(\omega^{2}-\omega_{\mathrm{\mathrm{cl:Q}}}^{2})$, $D[\omega]=(\omega
^{2}-\omega_{\mathrm{\mathrm{hd:B}}}^{2})(\omega^{2}-\omega_{\mathrm{
\mathrm{hd:Q}}}^{2})$, $E[\omega]=\omega(\omega^{2}-\omega
_{\mathrm{\mathrm{cl:Q2}}}^{2})$ and $F[\omega]=(\omega^{2}-\omega
_{\mathrm{\mathrm{hd:Q2}}}^{2})$. Each term represents two equations since
they contain real and imaginary parts. For an elongated cigar-shape trap it is
possible to write the frequencies in the form of Eq.\,(\ref{equ:frequ}) with
rescaled relaxation time $\tilde{\tau}=\left(  \omega_{\mathrm{cl:B}}
^{2}/\omega_{\mathrm{hd:B}}^{2}\right)  \tau$, which reaches the value
$\tilde{\tau}=6/5\,\tau$ in the harmonic limit. The numerically calculated
results of temperature induced shift, based on Eq.\,(\ref{eqn.TrapField}) in
the cross-over regime is represented by the grey sector in
Fig.\,\ref{fig:QUADFrequ}.

A comparison with experiment is beyond the scope of this paper because, after
scaling to $\omega_{Q}$, the two limiting cases are spaced by only $\sim
\,1\%$. Therefore, not only $\tilde{\omega}_{Q}$ has to be determined to an
accuracy much better than $1\%$, but also the scaling parameter $\omega_{Q}$.
In the limiting cases the latter is fully determined by the trap frequency
(i.e. $\omega_{\mathrm{cl}}=2\omega_{z}$ and $\omega_{\mathrm{hd}}
\approx1.55\omega_{z}$). However, in the crossover region knowledge of
$\tilde{\tau}$ to much better than $1\%$ is required to calculate $\omega_{Q}$
from Eq.\,(\ref{equ:FrequCrossOver}) to adequate precision.

\section{Results and discussion}

\begin{figure}[pb]
\includegraphics[width=80mm]{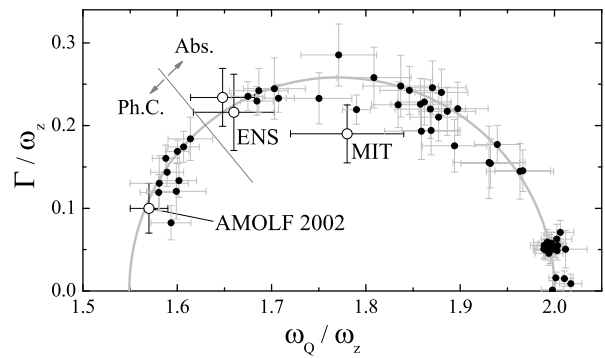}\caption{Damping (raw
data) versus frequency (extrapolated) for the quadrupole mode. Open circles:
data acquired at MIT \cite{Ketterle 1999}, ENS \cite{Leduc} and AMOLF
\cite{Shvarchuck 2002}. The solid line is corresponds to the crossover
expression Eq.\,(\ref{equ:frequ}). The data left of the straight line are
obtained with Phase Contrast (Ph.C.) imaging, those on the right with
Absorption (Abs.) imaging. The error bars represent a 95\% confidence interval
of the fit.}%
\label{fig.crossover}%
\end{figure}

\begin{figure}[pt]
\includegraphics[width=80mm]{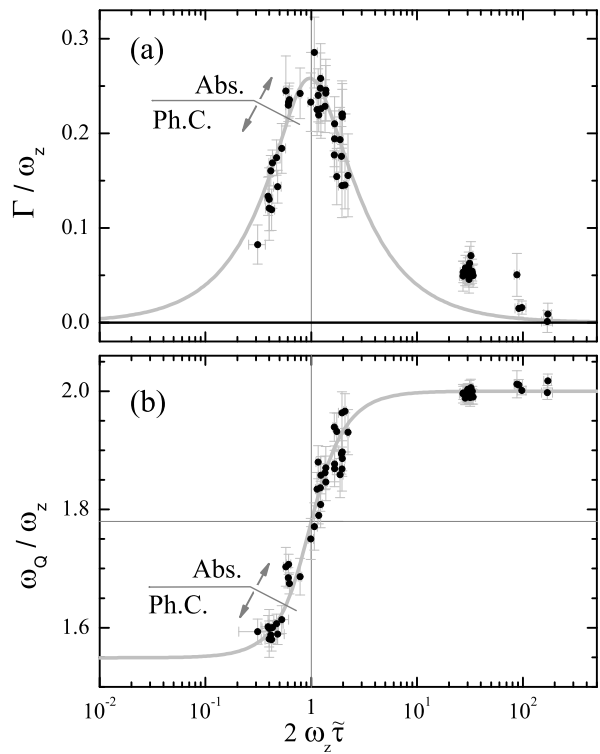}\caption{Damping (raw data)
(a) and frequency (extrapolated to harmonic limit) (b) of the quadrupole mode
versus scaled relaxation time. The solid line in figure a) corresponds to the
imaginary part $\left(  \Gamma=-\omega^{\prime\prime}\right)  $ of the
solution of Eq.\,(\ref{equ:frequ}); in figure b) to the real part $\left(
\omega_{Q} =\omega^{\prime}\right)  $. The crosshair marks the location of the
crossover point. The vertical error bars are identical to
Fig.\,\ref{fig.crossover}, the horizontal ones represent the standard
deviations.}%
\label{fig.DampingAndFrequency}%
\end{figure}

We took all our data with the same trap parameters and the same excitation
procedure, but at various temperatures. Starting the evaporation with a large
atom number and using `tight' trapping parameters we could reach high
densities and thus study the full crossover. However, this choice for a
`tight' trap made us sensitive for anharmonic shifts as discussed in section
\ref{sec:nonlinear} \cite{Note Frequency shifts}. Therefore, we extrapolate
all frequency data to the zero-temperature limit ($\tilde{\omega}%
_{Q}\rightarrow\omega_{Q}$) using the dotted curve in
Fig.\,\ref{fig:QUADFrequ}. This yields the best estimate for the value in the
harmonic limit of our potential. The correction curve is based on the
temperature dependence observed for our data in the collisionless regime
$\left(  2\omega_{z}\tilde{\tau}>10\text{, see
Fig.\,\ref{fig.DampingAndFrequency}}\right)  $, where we may presume
$\omega_{Q}=2\omega_{z}$. In this way systematic deviations of our results
from the curves in Fig.\,\ref{fig.crossover} and
Fig.\,\ref{fig.DampingAndFrequency}b were substantially reduced.

In Fig.\,\ref{fig.crossover} we plot the observed, scaled damping rates
$\Gamma/\omega_{z}$ versus the extrapolated quadrupole frequencies normalized
to the axial trap frequency, $\omega_{Q}/\omega_{z}$. The drawn curve
corresponds to the crossover expression, Eq.\,(\ref{equ:frequ}) with
$\omega_{\mathrm{cl}}=2\omega_{z}$ and $\omega_{\mathrm{hd}} =1.55\omega_{z}$.
Plots of the same experimental data and the exact solutions of
Eq.\,(\ref{equ:frequ}) separately against $2\omega_{z}\tilde{\tau}$ are given
in Fig.\,\ref{fig.DampingAndFrequency}a for $\Gamma/\omega_{z}$ and
Fig.\,\ref{fig.DampingAndFrequency}b for $\omega_{Q}/\omega_{z}$.

From the damping results (Fig.\,\ref{fig.DampingAndFrequency}a) we obtain
$2\omega_{z}\tilde{\tau}_{0}=1.0(1)$ for the experimental value of the
crossover point. Given the $30\%$ absolute accuracy of our density
determination this agreement is fortuitously good (see
Sec.\ref{sec:Background}).

The determination of the crossover point from the frequency crossover behavior
is less straightforward, because errors in the temperature determination add
to the error in the extrapolated frequency $\omega_{Q}$. Further, as the
frequency corrections are all positive, they affect the determination of
$2\omega_{z}\tilde{\tau}_{0}$ from Fig.\,\ref{fig.DampingAndFrequency}b. For
the crossover region $\left(  0.1\lesssim2\omega_{z}\tilde{\tau}%
\lesssim10\right)  $ the average applied frequency correction was
$\delta\omega_{Q}/\omega_{z}=2.6\%,$ which changes the experimental value for
$2\omega_{z}\tilde{\tau}_{0}$ by a factor $0.85$ to yield $2\omega_{z}
\tilde{\tau}_{0}=1.0(1)$ \cite{Note ErrorFrequencyShift}. This value coincides
with the result obtained from the damping data and shows that our results are
self-consistent. 

In the collisionless regime anharmonicities can give rise to dephasing induced
damping. These effects were not corrected for as they do not affect to leading
order the determination of the crossover point. Here we briefly comment on
these effects. Roughly, one may argue that for a given anharmonic spread
$\delta\omega_{Q}$ in frequencies the dephasing time $\delta t$ will be given
by $\delta\omega_{Q}\delta t\approx2\pi$. Hence, the dephasing related damping
rate is $\delta\Gamma=2\pi/\delta t\approx\delta\omega_{Q}$. The cluster of
data points at $2\omega_{z}\tilde{\tau}\approx30$ in
Fig.\,\ref{fig.DampingAndFrequency}a best illustrates the significance of the
correction as they were taken at the highest temperature ($9$ $\mu$K). For
these points the anharmonic frequency shift $\delta\omega_{Q}$ is $\sim3.5\%$
(see Fig.\,\ref{fig:QUADFrequ}). With $\delta\Gamma/\omega_{z} \simeq
\delta\omega_{Q}/\omega_{z}=0.035$ this suggests that the anomalously high
damping rates observed for these data points
(Fig.\,\ref{fig.DampingAndFrequency}a) may be entirely attributed to dephasing
effects. Near the crossover point the collisional damping is much faster and
dephasing corrections may be neglected $\delta\Gamma/\Gamma\simeq\left(
1/2\right)  \left(  \delta\omega_{Q}/\Gamma\right)  ^{2}\approx10^{-3}$.

We also verified that our shot-to-shot variations in the density have a
negligible effect on the measured damping rate. The frequency shifts fastest
at the density of the crossover point, where $\delta\left(  \omega_{Q}
/\omega_{z}\right)  /\delta\left(  \omega_{\mathrm{cl}}\tilde{\tau}\right)
=(2/\pi)\left(  \omega_{\mathrm{cl}}-\omega_{\mathrm{hd}}\right)  /\omega
_{z}\approx0.3$ as follows directly by taking the first derivative of
Eq.\,(\ref{equ:FrequCrossOver}) with respect to $\tilde{\tau}$. As
$\tilde{\tau}$ scales inversely proportional to the central density, a $1\%$
variation in atom number results (at constant temperature) in a $\sim0.3\%$
variation of the frequency, which is much smaller than to the one considered
above and therefore also negligible.

\section{Summary and Conclusions}

With Fig.\,\ref{fig.crossover} and Fig.\,\ref{fig.DampingAndFrequency} we
obtain good agreement between experiment and the crossover theory. The
frequency shifts down from $2\omega_{z}$ in the collisionless regime to
$1.55\omega_{z}$ for collisionally hydrodynamic clouds, with $\omega_{z}$ the
axial frequency of our trap. Most of the shift occurs over a narrow range of
densities around the crossover point. The damping rate peaks over the same
range of densities. The determinations of the crossover point from the
frequency and the damping behavior agree within $10\%$, $2\omega_{z}
\tilde{\tau}_{0}=1.0(1)$. The agreement with theory is limited by a $30\%$
absolute uncertainty in density. Further, we present a theory and experimental
evidence for anharmonic frequency shifts. The theory allows numerical
evaluation for potentials with known first and second spatial derivatives. We
show that for elongated Ioffe-Pritchard traps knowledge of the central field
$B_{0}$ suffices to calculate the leading anharmonic shifts with simple
analytic expressions.

\begin{acknowledgments}
The authors wish to thank Steve Gensemer, Mikhail Baranov and
J\'{e}r\'{e}mie L\'{e}onard (JL) for valuable discussions and JL
also for providing data from his thesis. JTMW benefitted from
participation in the program on quantum gases of the Kavli
Institute for Theoretical Physics (KITP) in Santa Barbara. This
work is part of the research program on cold atoms of the
Stichting voor Fundamenteel Onderzoek der Materie (FOM), which is
financially supported by the Nederlandse Organisatie voor
Wetenschappelijk Onderzoek (NWO).
\end{acknowledgments}


\begin{thebibliography}{99}
\bibitem{Note}Because the flow is irrotational shear viscosity plays no role
and viscous damping is absent. Because the flow is divergence-free the
temperature is constant.
\bibitem{Laser Depletion}The depletion is done with our detection laser. For
absorption imaging we typically use $5\%$ of the full detection intensity
($2.4$ mW) for typically $10$ ms at a detuning of $32$ MHz red to the
$(5S_{1/2},F=2)\leftrightarrow(5P_{3/2},F=3)$ transition ($D_{2}$-line). For
phase-contrast imaging we use $>80\%$ of the full detection intensity at a
detuning of $-3$ GHz for a duration of up to $100$ ms.
\bibitem{Note on Gravity}The gravitational force $mg$ is small as compared to
the gradient of the radial trapping field, $mg/\mu_{B}\alpha\approx0.04$.
Therefore gravity corrections may be neglected to lowest order in the expansion.
\bibitem{Detection}The lasers for phase-contrast and absorption imaging in
our experiment are separate devices, the first operating in a region of
$\pm\,3\,$GHz (no online control), the second from $-32$ to $+15\,$MHz (online
control) with respect to the $D_{2}$-line. It was not possible to use both
detection methods during the same experimental run.
\bibitem{Note on detection}Detection is done with a $40\,\mu$s flash of
linearly polarized light and for absorption imaging detunings chosen to keep
the peak optical density at $\sim2$. The polarization and propagation vectors
are both orthogonal to the trap axis. The intensity is $2$ mW/cm$^{2}$. The
effects of saturation and optical pumping arise only at small detunings and
are corrected for to first order where necessary.
\bibitem{Note 2D versus 3D fitting}We note that $L_{e}$ is obtained from the
column density and does not coincide with the effective length of the cloud.
In contrast to harmonic traps, in anharmonic traps the value obtained for the
Gaussian length of a thermal cloud by fitting a 2D Gaussian to the column
density differs in general from the value obtained by fitting a 3D Gaussian to
the full density distribution of the same cloud.
\bibitem{Repeating}Further investigation of this anharmonicity or
reacquisition of the quadrupole oscillation data was not possible, since the
experimental apparatus was disassembled prior to the completion of data analysis.
\bibitem{Note Frequency shifts}In the collisionless limit frequency shifts as
small as $1\%$ are significant (see Fig.\,\ref{fig.crossover}).
For our trap anharmonic shifts of that size are expected for
$T\gtrsim2\,\mu$K.
\end{thebibliography}
\end{document}